\documentclass[widetext,amsmath,amssymb,twocolumn]{revtex4}%twocolumn
\usepackage{graphicx}% Include figure files
\usepackage{dcolumn}% Align table columns on decimal point
\usepackage{bm}% bold math
\usepackage{epstopdf}
\begin{document}
\title{On the problem of boundaries and scaling  for urban street networks}
\author{A. Paolo Masucci}
\altaffiliation{Centre for Advanced Spatial Analysis, University College of London, 90 Tottenham Court Road, London W1N 6TR, UK }
\author{Elsa Arcaute}
\altaffiliation{Centre for Advanced Spatial Analysis, University College of London, 90 Tottenham Court Road, London W1N 6TR, UK }
\author{Erez Hatna}
\altaffiliation{Center for Advanced Modeling, The Johns Hopkins University, USA}
\author{Kiril Stanilov}
\altaffiliation{Centre to Sustainable Infrastructure and Construction, University of Cambridge, Cambridge CB2 1TN, UK}
\altaffiliation{Centre for Advanced Spatial Analysis, University College of London, 90 Tottenham Court Road, London W1N 6TR, UK }
\author{ Michael Batty}
\altaffiliation{Centre for Advanced Spatial Analysis, University College of London, 90 Tottenham Court Road, London W1N 6TR, UK }
\date{\today}
%\pacs{ 89.75.-k, 89.65.Lm, 89.75.Fb, 89.75.Da}
%87.23.Kg Dynamics of evolution
%\PACS{%02.50.Ey Stochastic processes
%{02.50.Le}{Decision theory and game theory} \and
%87.23.-n Ecology and evolution
%{87.23.Ge}{Dynamics of social systems} \and
%87.23.Kg Dynamics of evolution
%89.65.-s Social and economic systems
%89.75.-k Complex systems
%89.75.Kd Patterns
%47.54.-r Pattern selection; pattern formation [GRANULAR] (see also
%{82.40.Ck}{Pattern formation in reactions with diffusion, flow and heat transfer} \and
%{87.23.Cc}{Population dynamics and ecological pattern formation}
% in Physical chemistry and chemical physics; 87.18.Hf Spatiotemporal pattern formation in cellular populations in Biological and medical physics)
%\keywords{Suggested keywords} % Use showkeys class option if keyword
                               % display desired

\begin{abstract}
Urban morphology has presented significant intellectual challenges to mathematicians and physicists ever since the eighteenth century, when Euler first explored the famous  K{\"o}nigsberg bridges problem.
Many important regularities and scaling laws have been observed in urban studies, including Zipf's law and Gibrat's law, rendering cities attractive systems for analysis within statistical physics. 
Nevertheless, a broad consensus on how cities and their boundaries are defined is still lacking.
Applying an elementary clustering technique to the street intersection space, we show that growth curves for the maximum cluster size of the largest cities in the UK and  in California collapse to a single curve, namely the logistic.
Subsequently, by introducing the concept of the condensation threshold,  we  show that natural boundaries of cities can be well defined in a universal way. 
This allows us to study and discuss systematically some of the regularities that are present in cities. 
We show that some scaling laws present consistent behaviour in space and time, thus suggesting the presence of common principles at the basis of the evolution of urban systems.
\end{abstract}
\maketitle
%%%%%%%%%%%%%%%%%%%%%%%%%%%%%%%%%%%%%%%%%%%%%%%%%%%%%%%%%%%%%%%%
\medskip
Since the middle of the twentieth century universal properties of cities have been identified, including  Zipf’s and Gibrat’s laws  \cite{zipf,gibrat}.
City size has been measured most commonly in terms of built area or population since Zipf's seminal book \cite{zipf} notwithstanding that most of the time city boundaries have been defined in terms of often arbitrary, fixed administrative boundaries.

Many different techniques to define cities have been suggested based on the analysis of urban growth \cite{murcio2013second,rybski2013distance,frasco2014spatially}, and recently a method using demographic and commuting data has been proposed \cite{arcaute_Interface}.
Clustering techniques such as the City Clustering Algorithm (CCA)  have been applied, mostly to analyse satellite images and demographic data \cite{maksenat,rozengib,rozenzipf}, but  these are rarely parameter free. 
A method proposing a bottom up approach that does not rely on highly aggregated  census data or on the interpretation of remotely sensed images is needed.

When we define a city we have to keep in mind that built area and population are strongly correlated \cite{rozenzipf}, but these correlations, as we show in this paper, do not necessarily carry universal exponents.
The interpretation of the empirical outcomes using these definitions have to be therefore put into context according to the methodology employed.

As pointed out in \cite{arcaute_Interface}, a broad range of exponents based on different allometries inferred from urban studies \cite{westsublin,lammer} can be observed for different boundary definitions. 
This further supports the urgent need for an operational and context-free definition of the city.
It is somewhat astonishing that in spite of the large body of literature about cities, the very concept of  city remains in some ways obscure, hidden or assumed.

 \begin{figure} [t]
\begin{center}
	\includegraphics[width=0.4\textwidth]{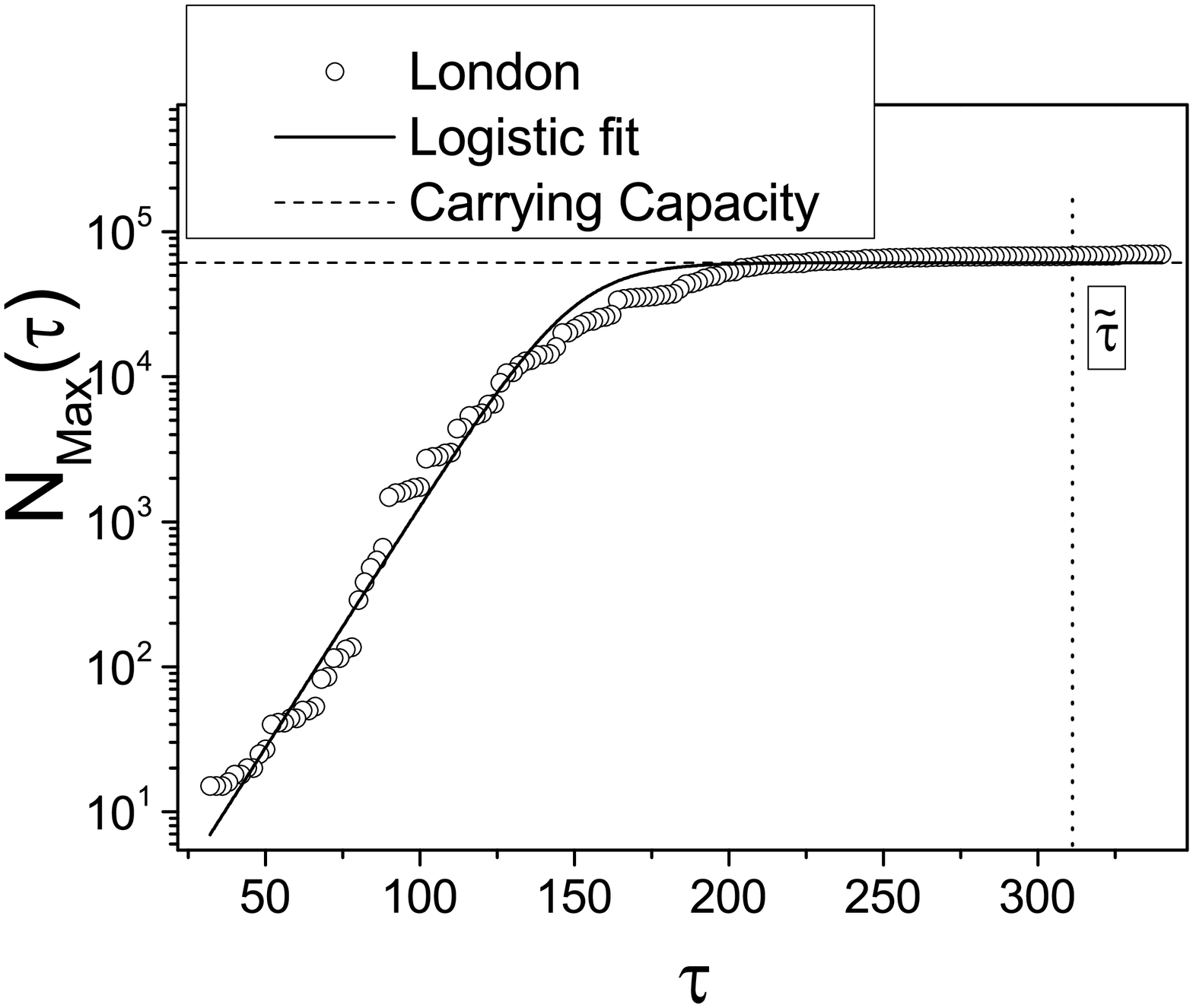}
	\includegraphics[width=0.35\textwidth]{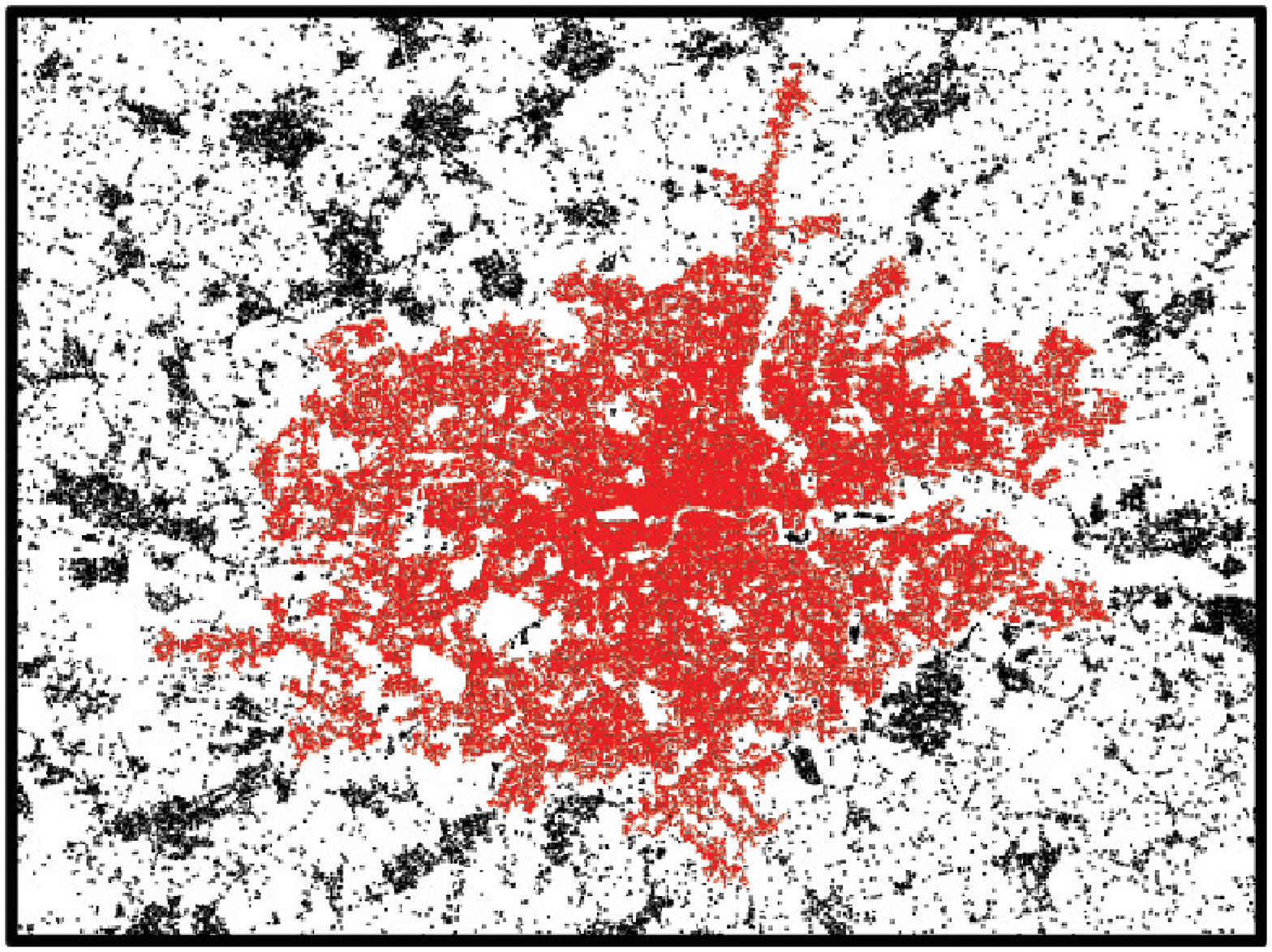}
	\caption{\textbf{Logistic growth for the maximum cluster size in a clustering process: the condensation threshold.} Top Panel: Maximum cluster size $N_{Max}(\tau)$ as a function of the threshold $\tau$ for Greater London on a semi-log plot. The solid line is the logistic function fit of Eq. \ref{eq1}. The dashed line represents the carrying capacity $C$, while the dotted line shows the condensation threshold $\tilde{\tau}$, defined as the threshold where $N_{Max}(\tau)=C$.
	Bottom Panel: the maximum cluster  (red) at the condensation threshold for the London.}
	\label{fig1}
\end{center} 
\end{figure}

In this paper we present some universal properties of cities which emerge when applying an elemental clustering technique to the  vertices and edges of street networks. 
We obtain a logistic growth curve from which the structural fringe of the city can be defined mathematically in a bottom-up approach.
This is achieved by obtaining the parameters at the point at which a \emph{condensation} phenomenon is observed as we will explain below.
The curves for all cities then collapse to a single curve, and city boundaries are hence defined in a universal way. Such universality in the spatial properties of cities prompts us to look at the spatial and temporal behaviour of important properties of urban street networks, and thus investigate whether some scaling laws could display a general behaviour.

\begin{figure*} [t]
\begin{center}
	\includegraphics[width=0.4\textwidth]{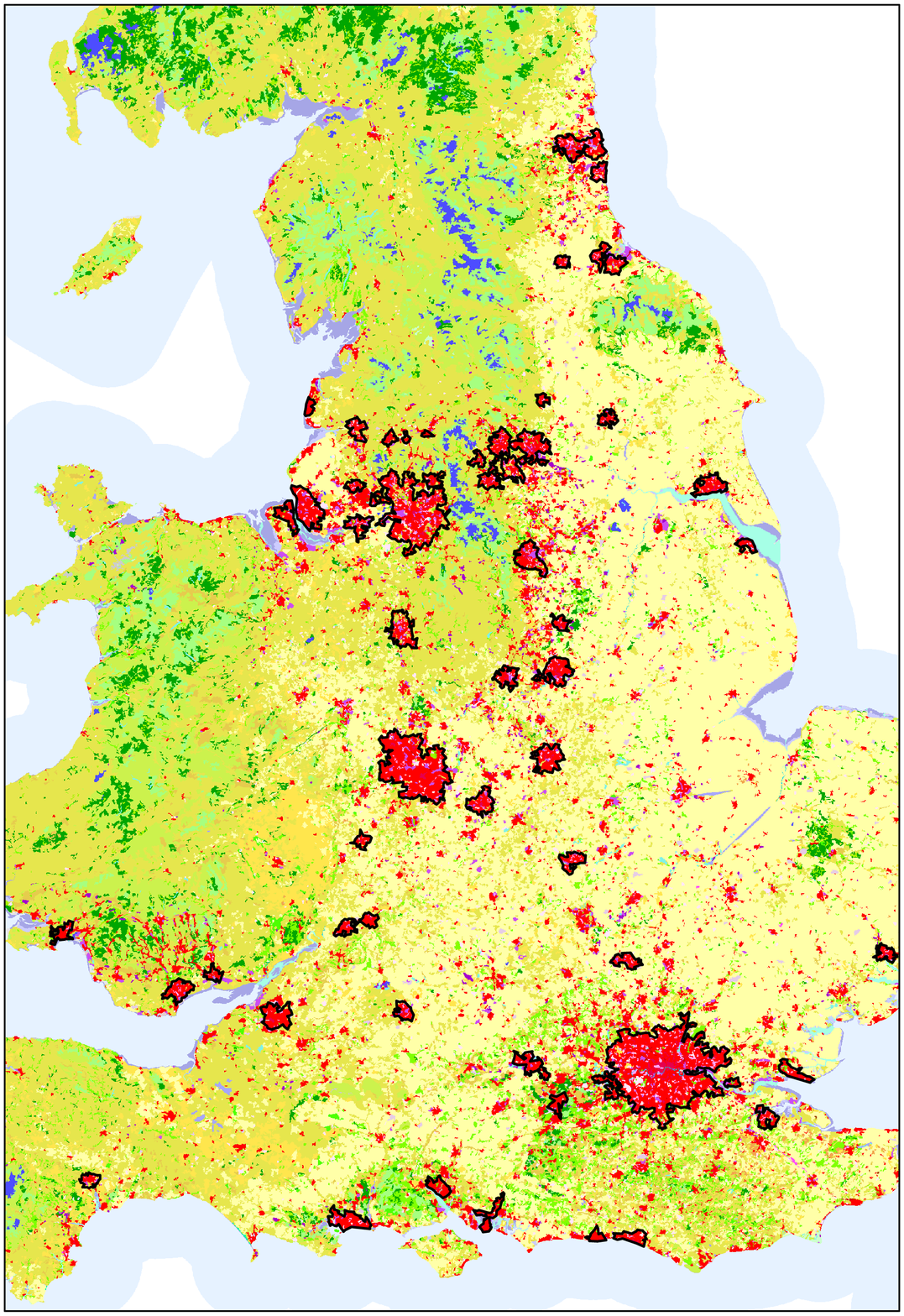}
	\includegraphics[width=0.4\textwidth]{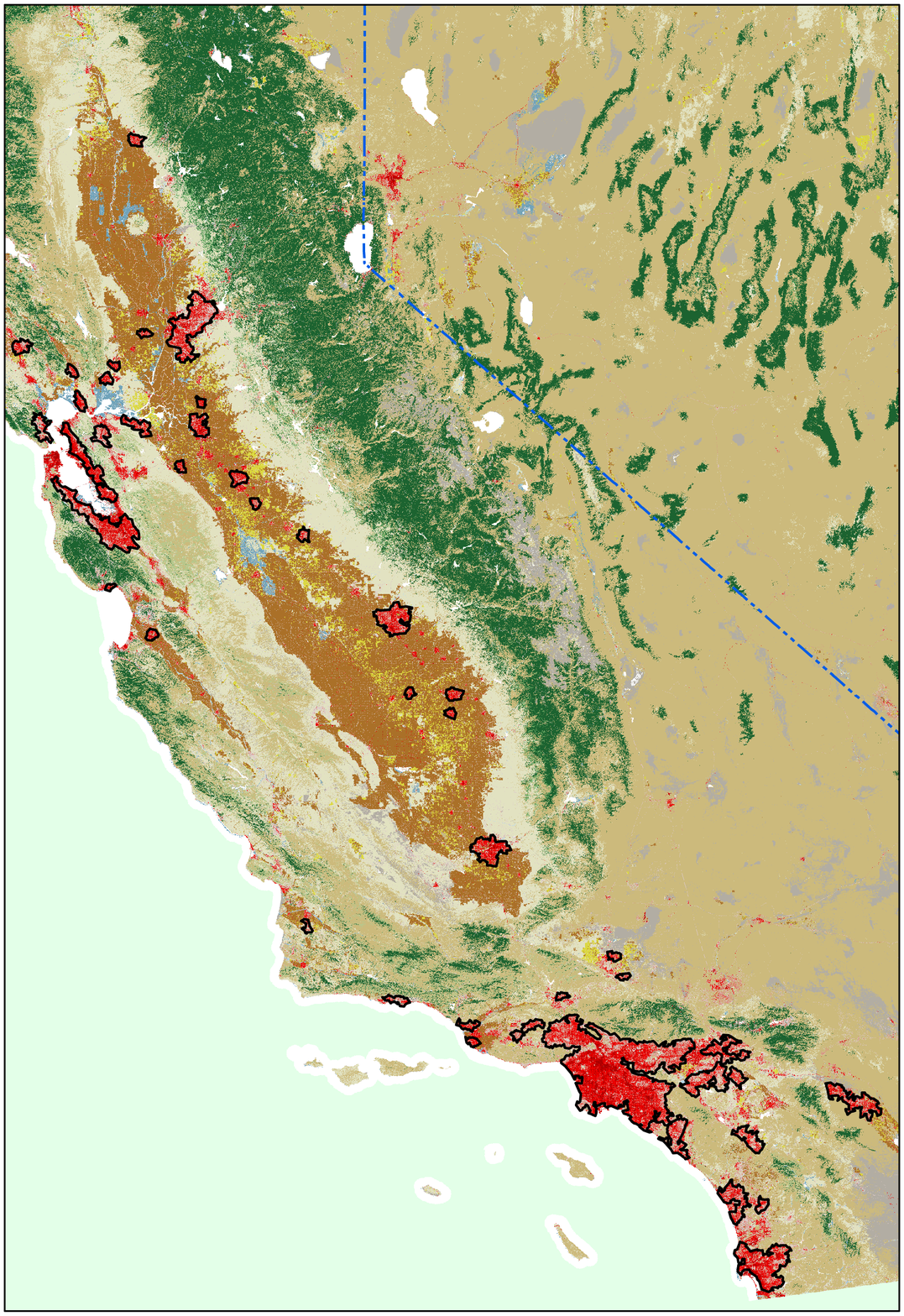}
	\caption{\textbf{The UK and California datasets with land-use satellite comparison} In the left panel a large portion from the Corine data set for the UK map representing a satellite image with land-use. In the right panel the California satellite land-use map. The red parts are identified as urban areas, while the black contours are the city condensation boundaries as defined in the text. Note that throughout this paper, we refer to towns and cities as being in the UK when strictly we are excluding those in Northern Ireland}
	\label{fig3}
\end{center} 
\end{figure*}

\section{Results}

A city is a complex organism, composed of many superimposing layers, such as transportation networks, the built environment, and different economic, social, and information flows \cite{mikecc,mikefract,barth2011}. 
Such layers are dynamical by nature, and give rise to generic patterns, such as fractal geometries \cite{mikecc,mikefract}.
Administrative boundaries overlook these aspects, and are not able to measure or record the dynamical aspects of cities in a consistent way across space. 

Between others, street networks provide a good representation to characterise the morphology of a city, where a street network is defined as that planar graph where the street intersections $N$ are the vertices and the street segments $E$ are the links. 
We will consider here street intersections as being a good proxy for the urbanization process.
Such a choice reduces the complexity of the problem to that of a spatial point pattern.
This has the value of simplicity. 
Moreover, it has been positively tested before \cite{masuccipo,masuccilond} with some correlations between the number of street intersections and built area for urban systems are shown in the  Supporting Information (SI), Sec. I-D.

\begin{figure*} [t] \begin{center}
	\includegraphics[width=0.48\textwidth]{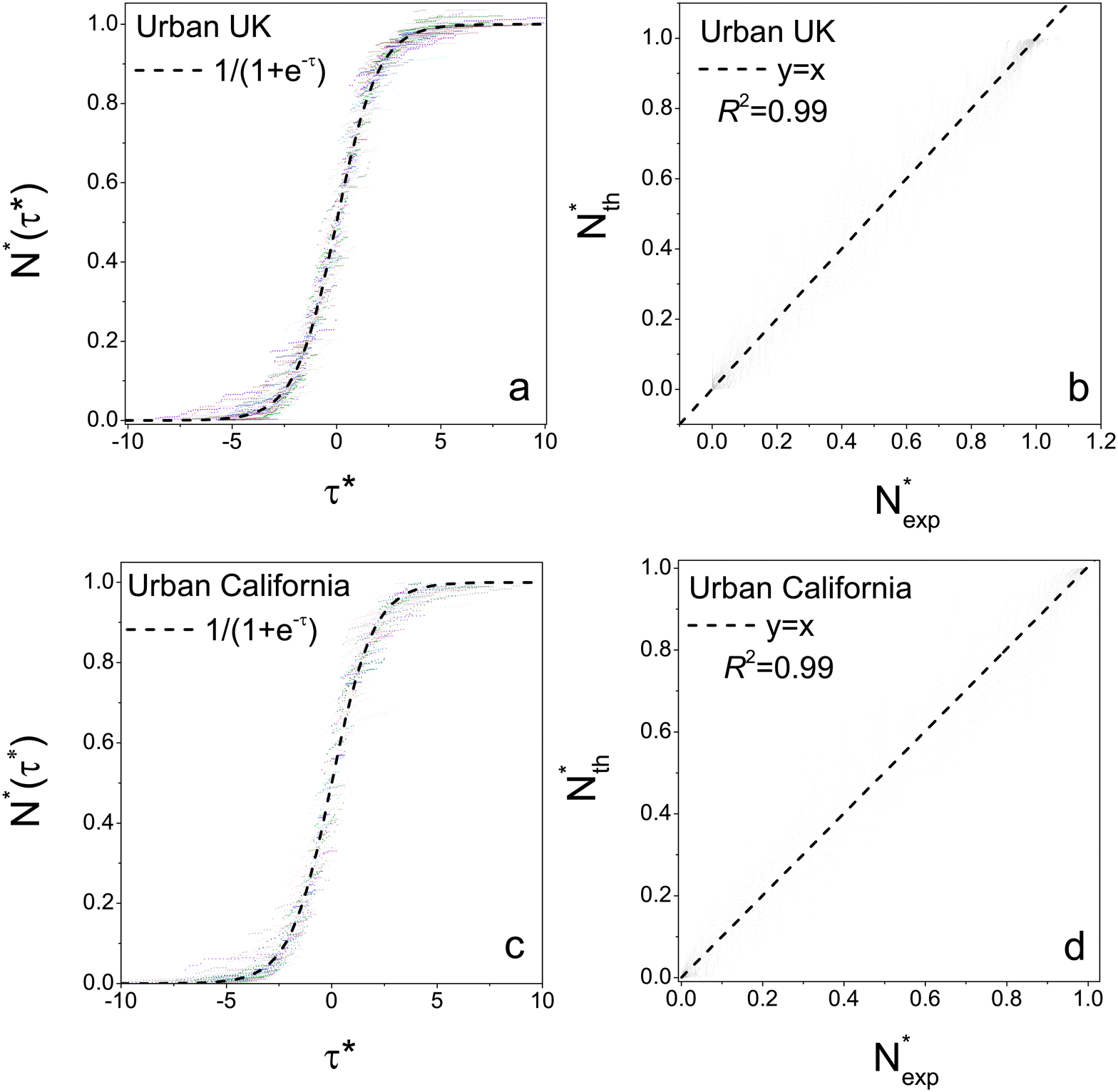}
	\includegraphics[width=0.48\textwidth]{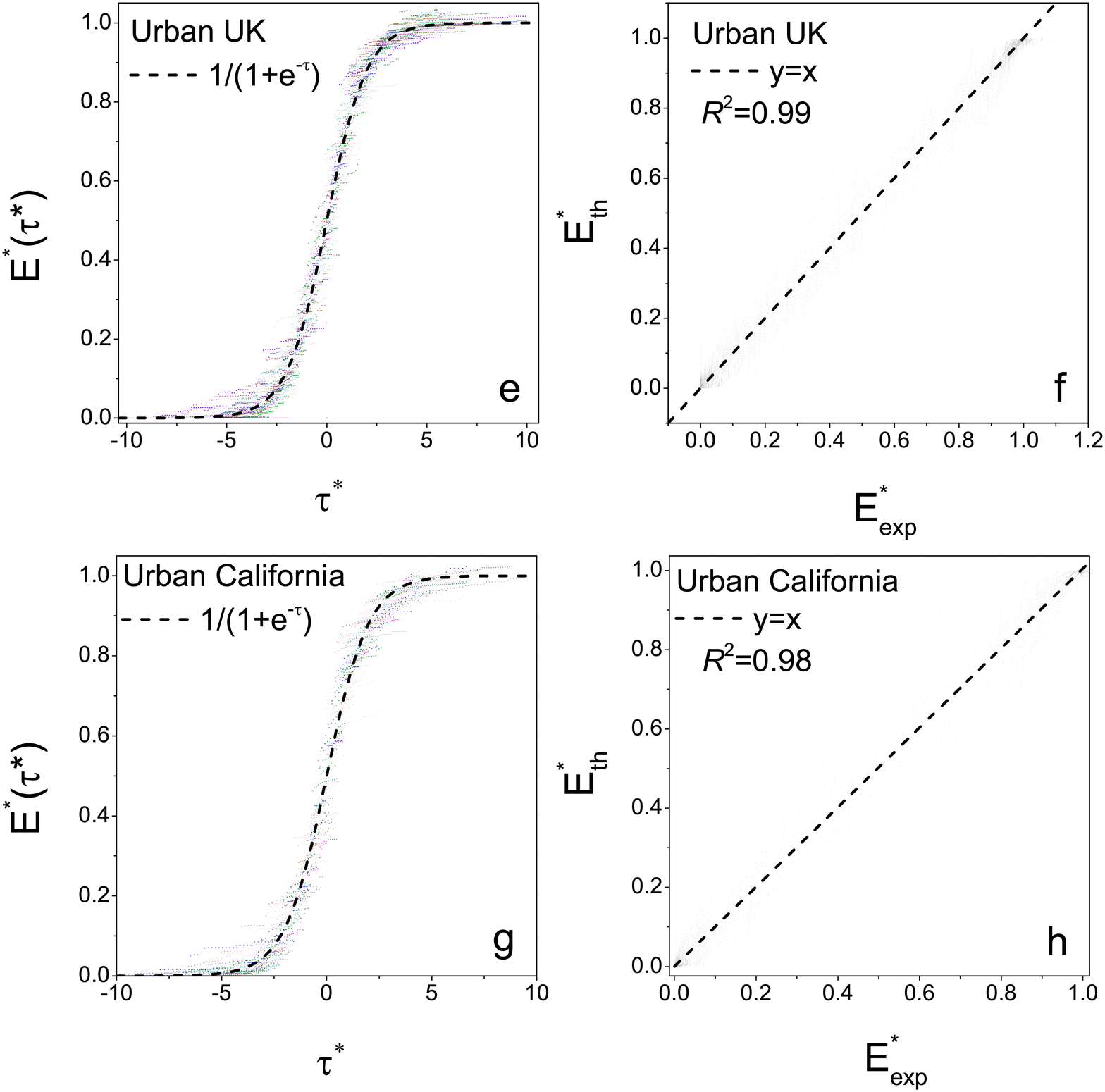}
	\caption{\textbf{Growth curve collapse for the cities in the UK and in California} Panels \textit{a, c}: rescaled maximum cluster size $N^*=N_{Max}(\tau)/C$ as a function of the rescaled threshold $\tau^*=r(\tau-\tau_0)$ for the largest 61 cities in the UK and for the largest 52 cities in California. The dashed curve is $1/(1+e^{-\tau})$.
	Panels\textit{ b, d}: in order to evaluate the goodness of the collapse of the curves in the figures in panels a and c, we plot in the horizontal axis the  $N^*_{exp}=N_{Max}(\tau)/C$ values for the cities in the UK (panel \textit{b}) and in California (panel \textit{d}) and in the vertical axis, the estimated value via the logistic function 	 $N^*_{th}=(1+e^{-\tau^*})^{-1}$.  
	 Then we calculate the $R^2$ value of the resulting points with the dashed curve $y=x$ and we find that $R^2>0.99$ for both UK and California cities. Panels \textit{e-h}: the same methodology as explained for panels \textit{a-d} is applied for the number of street segments $E(\tau)$ for the cities of the UK and California. In this case we find that the quality of the logistic collapse is given by $R^2>0.99$ for the UK and $R^2>0.98$ for California.}
	\label{fig2}
\end{center} 
\end{figure*}

Considering a spatial window large enough to contain a given city and using an elementary clustering technique \cite{BJ2011}, we consider two street intersections to belong to the same cluster if they have a distance below a given distance threshold $ \tau $, where $\tau$ is measured in meters. 
Increasing $\tau$ enlarges the size of the clusters, until eventually a giant component appears, which spans the entire street network. 

We measure the maximum cluster size $N_{Max}(\tau)$ in terms of number of intersections as a function of the increasing threshold $\tau$, and we find that for all the cities  $N_{Max}(\tau)$ grows exponentially  and eventually the growth slows down and the curve condensates to a certain value (see top panel Fig.1).
This behaviour has been positively tested for all the largest cities in the UK and in California, suggesting that the maximum cluster size behaviour for cities highlights universal properties of urban morphology (see SI Sec. II for more details).

\subsection{The condensation threshold}
The function defined by $N_{Max}(\tau)$, i.e., exponential growth followed by condensation, has the characteristics of the logistic growth function:
\begin{equation}\label{eq1}
N_{Max}(\tau)=\frac{C}{1+e^{-r(\tau-\tau_0)}}
\end{equation}
where $C$ is the carrying capacity, $r$ is the growth rate and $\tau_0$ is the inflection point \cite{Ver}.

Following Eq. \ref{eq1}, we show that for cities in the UK and in California, $N_{Max}(\tau)$ grows as $e^{r\tau}$ until the inflection point $\tau_0$, and after that it condensates at a constant value given by the carrying capacity $C$.
In order to do that,  given the transformation  $\{\tau^*=r(\tau-\tau_0),N^*=N_{Max}(\tau)/C\}$, we expect that all the measured curves would collapse to a single curve, namely $N^*(\tau^*)=1/(1+e^{-\tau^*})$.

We test this hypothesis  for the 61 largest  cities in the UK and for the 52 largest cities in California (see Fig. \ref{fig3} and SI, Sec. I). 
The results are shown in Fig. \ref{fig2}, and we can see that for both cases there is a very high correlation ($R^2>0.99$) for the quality of the collapse.
This correlation is maintained if the maximum cluster size is measured according to the number of street segments $E(\tau)$ instead of the number of intersections.
In this case we find that the  collapse is estimated with an $R^2>0.98$. 

\begin{figure} [t] \begin{center}
	\includegraphics[width=0.5\textwidth]{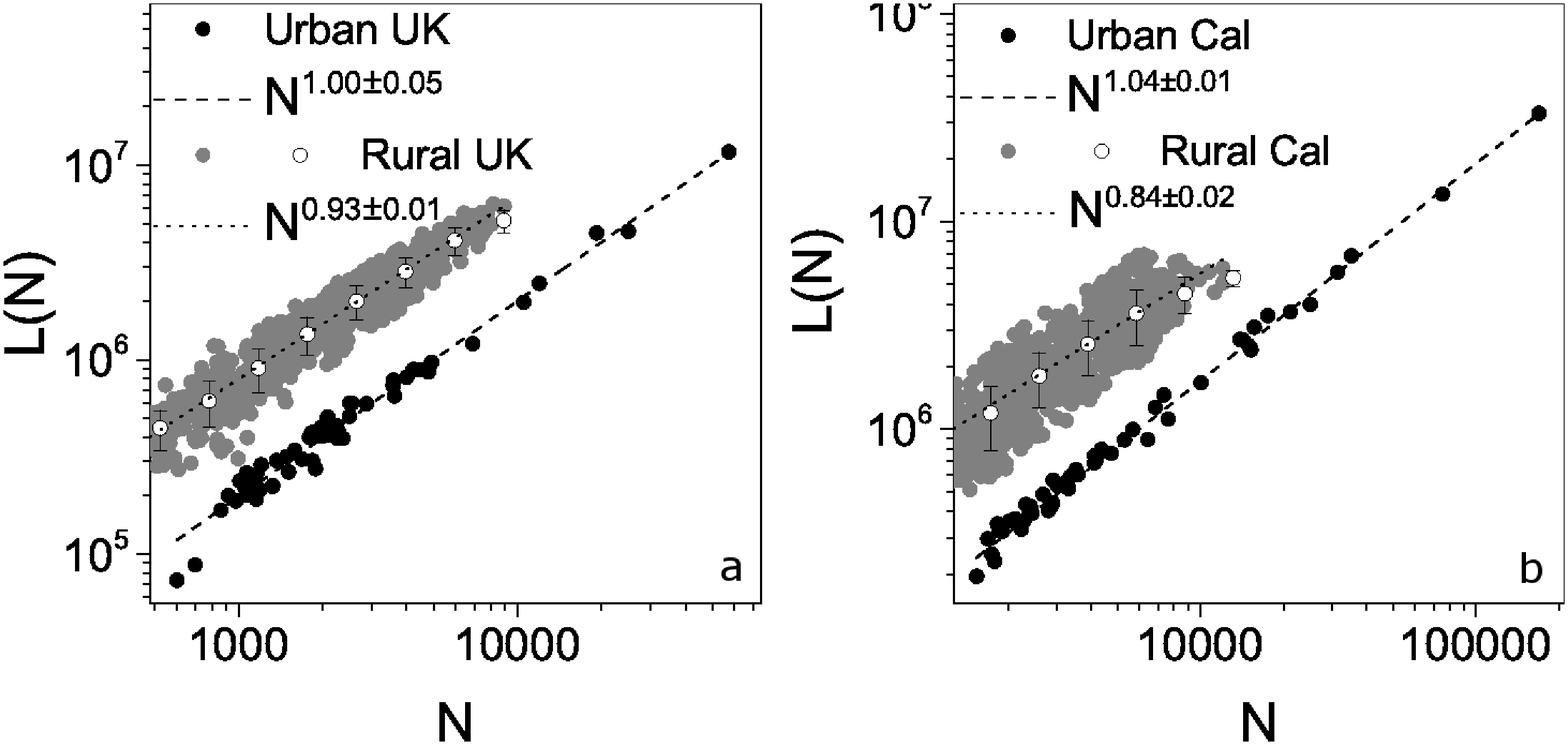}
	\includegraphics[width=0.5\textwidth]{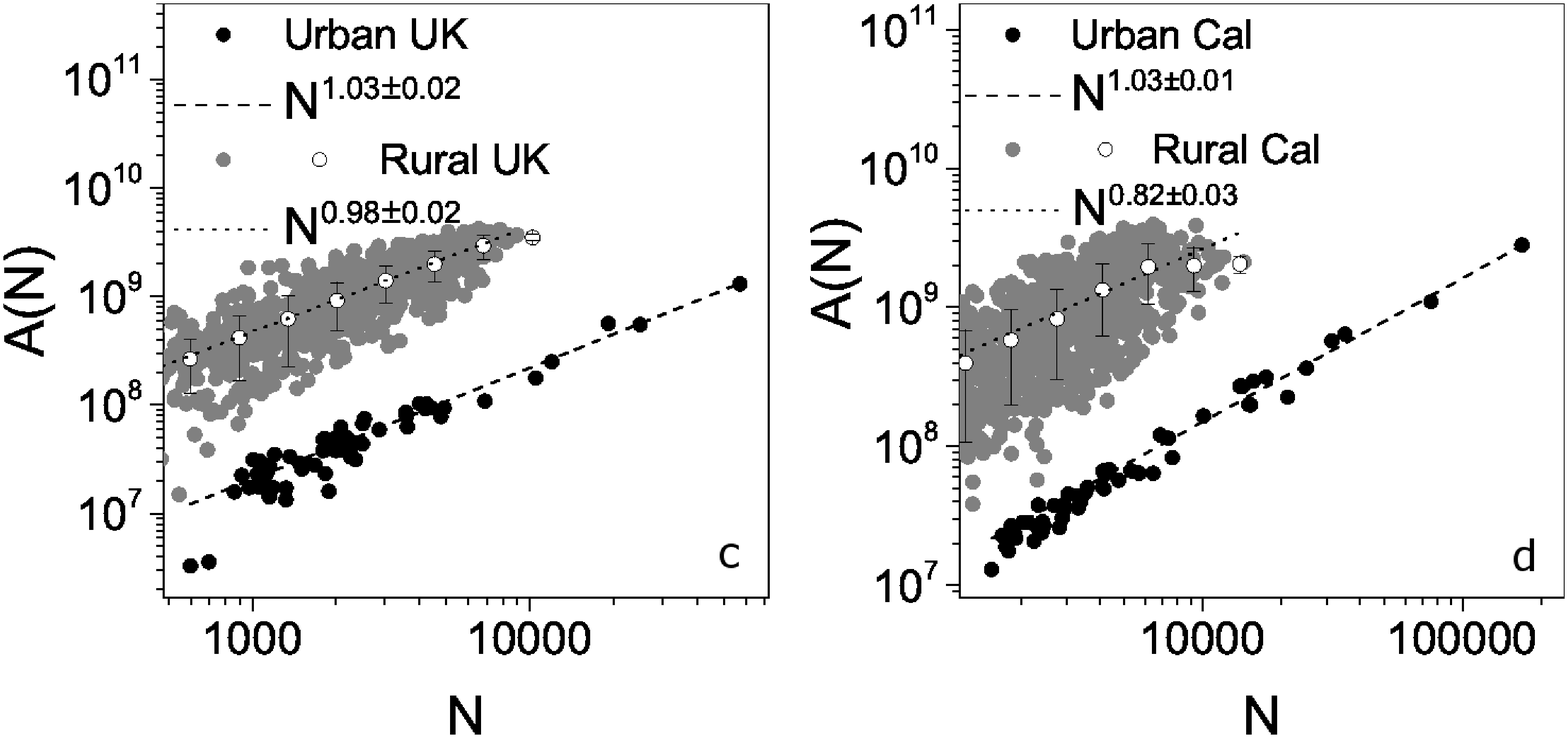}
	\includegraphics[width=0.5\textwidth]{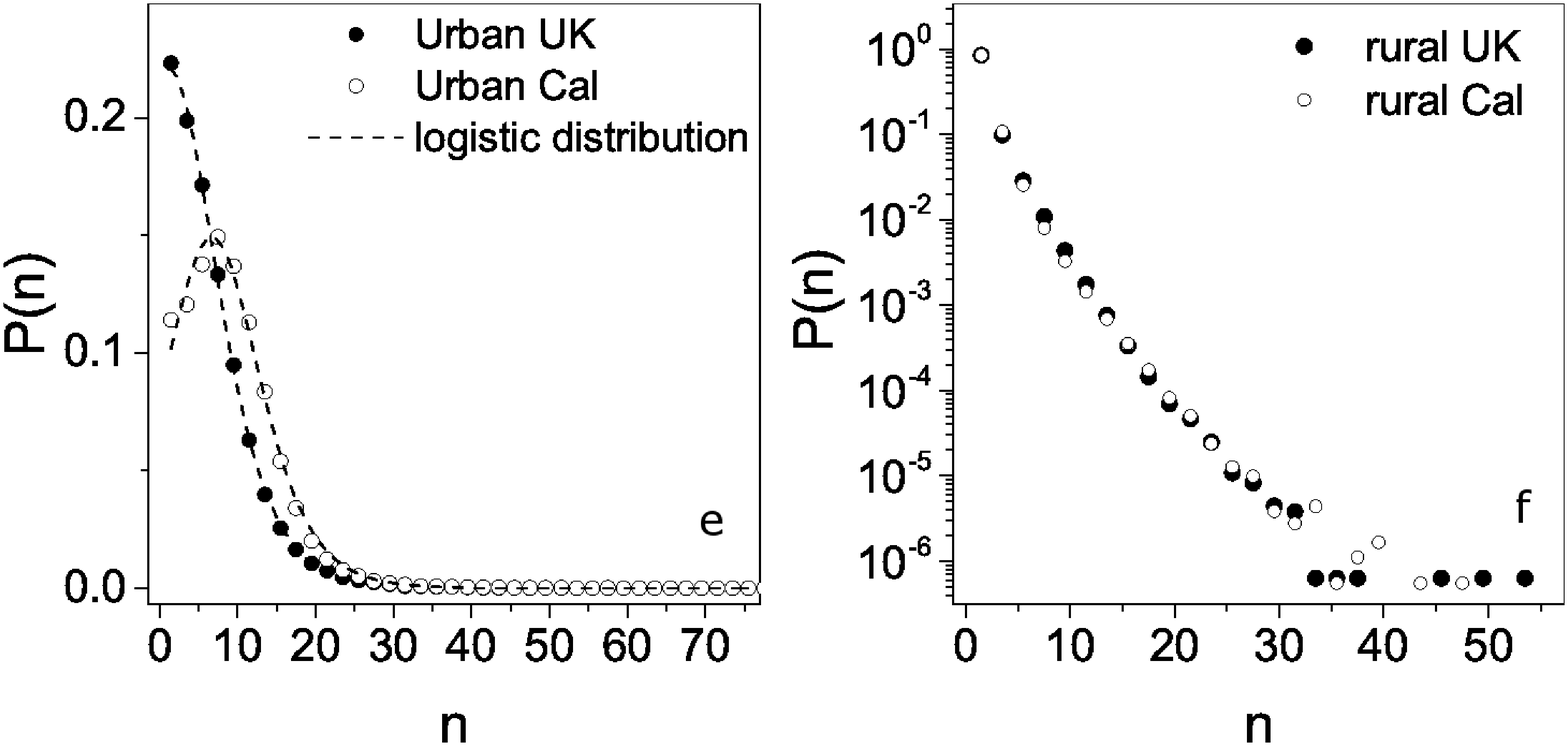}
	\caption{\textbf{Statistical properties of urban networks}  
 \textit{Panels a, b}: total length of the street network as a function of the number of  vertices $L(N)$  for the UK and California. 
   \textit{Panels c, d}: total area of the street networks as a function of the number of intersections $A(N)$ for the UK and California.
	 \textit{Panels e, f}: density distribution $P(n)$ for the number of intersections contained in a square grid lattice of 400 m. side for the UK and California. }
	\label{fig4}
\end{center} 
\end{figure}

These results indicate that the proposed clustering technique is able to capture generic properties of urban street networks.
In order to investigate this further, we look at how the logistic form of Eq.~1 is related to urban morphology and whether it allows us to define in a rigorous way the boundaries of a city.

As the logistic function is associated to the Verhulst model \cite{Ver}, it is interesting to understand how the carrying capacity $C$, always referring to a reservoir in the system, could be associated to our clustering approach.
To understand this, we notice that the largest cluster grows in the area where the intersection density is large, i.e., the urban area (See Fig. 1 as a visual reference).
The existence of a condensation phase shows that there exits an abrupt transition between the urban area and the rural area, where the intersection density consistently drops. 
Hence, the reservoir could be interpreted as the set of intersections belonging to the urban network which are consumed while the maximum cluster grows, and then the carrying capacity represents the city size in terms of street intersections.

Following  the clustering analysis introduced above, when $\tau$ grows after the logistic condensation phase, $N_{Max}(\tau)$ starts to grow again (see Fig.1). This is because after the maximum cluster reaches the condensation phase, as $\tau$ grows rural intersections and small towns close by  get absorbed by the maximum cluster. 
In such a way $N_{Max}(\tau)$ exceeds the carrying capacity $C$.
 
We  define the \textit{city condensation threshold} $\tilde{\tau}$, as the threshold where the measured maximum cluster size $N_{Max}(\tau)$ intersects the carrying capacity of the fitted logistic function, i.e. $\tilde{\tau}\equiv \tau:N_{Max}(\tau)=C$.
The city is so defined as the maximum cluster  at the city condensation threshold, as we show in Fig. 1 for London.
In order to investigate whether the city boundaries obtained in this way bear any resemblance with the urbanised space, we overlap the given contours with land-use satellite images.
Fig. \ref{fig3} demonstrates clearly  that the city boundaries as defined via the condensation threshold delimit the so called \textit{urban fringe}, i.e. the spatial pattern related to the city's expansion.

\subsection{Space and time scaling relations}
%Ergodicity is a fundamental property for statistical physics phenomena.
%It describes a system whose average measures in space and time are equivalent over all the system's states. 
%Therefore, investigating whether the urban system is ergodic or not is of great relevance. 
%On the one hand, if statistical properties of certain urban forms are found to be universal, and the system is ergodic, we could infer the historical evolution of cities from the spatial distribution of modern cities.
%On the other, if urban systems do not present universal spatial features, we advance towards an understanding that the evolution of cities is highly constrained  spatially, culturally and contextually. 

In the following section, we try to understand  the meaning of  different allometries that are usually found in urban studies and  we examine them in spatial and temporal terms.
To pursue this, we analyse a few simple global statistical properties of the spatial networks: the network total street length $L(N)$, measured in meters, which is the sum of the lengths of the street segments for a given network;  the network area $A(N)$, measured in square meters, which is the area embedded by a given street network; the street intersection density $P(n)$, obtained by imposing a 400 meters side square grid on the top of the street network, and counting the number $n$ of intersections falling in each cell \footnote{The size of the cell is somehow arbitrary. In France for example, administrative urban boundaries are set according to a maximum of 200m separation threshold between buildings. In a highly dense urban system such as the UK, the choice of 400m seems to be a reasonable scale to allow that each square contains a fair amount of intersections}. 
These quantities are quite sensitive to the structure of the network and some of them have been considered in different studies \cite{lammer, masuccipo,masuccilond,barthelemy2013self,bartprl}.

The following analysis  shows that urban street networks, as defined via the condensation threshold,  display statistical properties which are consistently different from the statistical properties of \textit{rural} street networks \footnote{In order to define rural street networks, we delete from the maps all the cities defined by the condensation threshold and then we sample from the resulting maps 1000 random portions of street network from each map (see SI, Sec. I).}. 
Moreover, we show that the allometric exponents obtained for the above mentioned properties are compatible for cities in the UK and for cities in California.
Remarkably, we find that these exponents are compatible with the ones found for the growth of London during the last two centuries.  

The  network total street length $L(N)$ is a global quantity characterizing the nature of the underlying network.
We can write that $L=E\langle l\rangle$, where $E$ is the number of street segments and $\langle l\rangle$ is the average length of a street segment, if such a quantity can be well defined. 
Then, considering that the average degree of the network can be written as $\langle k\rangle=2E/N$, we have $L=\langle l\rangle\langle k\rangle N/2$, where the density distribution for both $l$ and $k$ have finite mean and variance.

\begin{figure} [hb]
\begin{center}
	\includegraphics[width=0.48\textwidth]{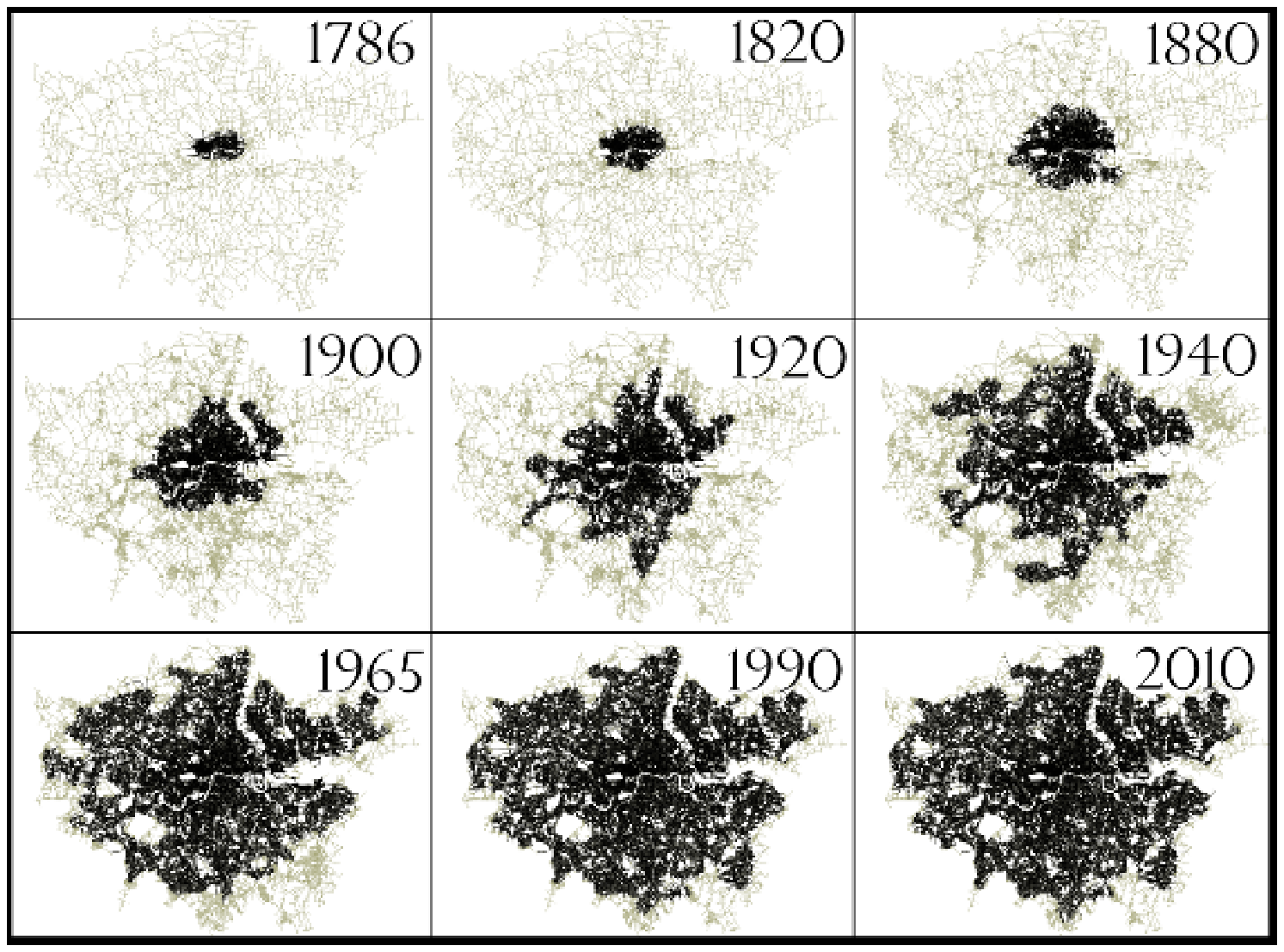}
	\caption{\textbf{Historical London} Street intersections of the city cores of London as defined by the condensation threshold from 1786 to 2010.}
	\label{fig5}
\end{center} 
\end{figure}
\begin{figure} [t]
\begin{center}
	\includegraphics[width=0.5\textwidth]{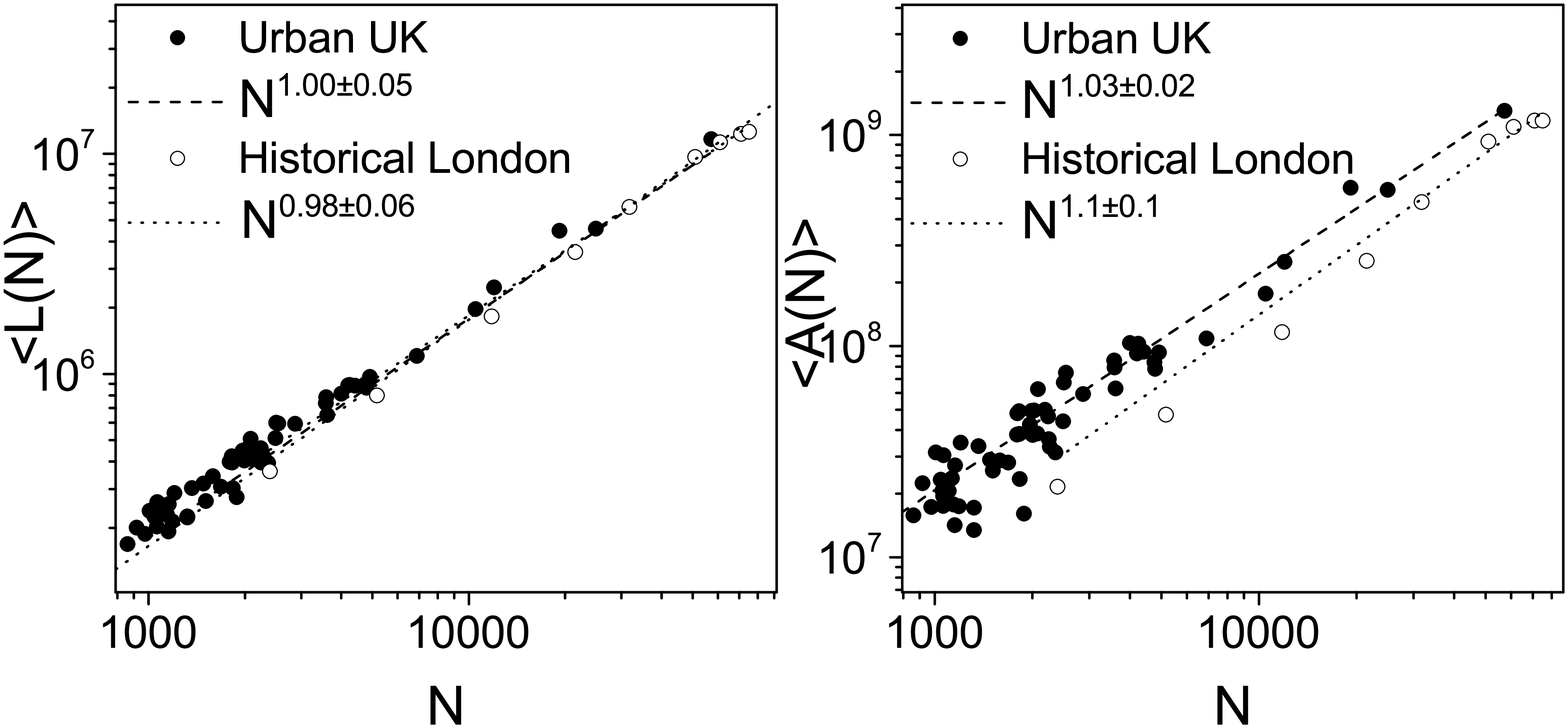}
	\caption{\textbf{Space-Time Analysis}  Left panel: total length of the street network $L(N)$ for the historical London dataset compared to the actual UK urban street networks. Right panel: area of the street network $A(N)$ for the historical London dataset compared to the actual UK urban street networks.}
	\label{fig6}
\end{center} 
\end{figure}

We find (see Fig. \ref{fig4}, panel \textit{a})  that for  cities in the UK the behaviour of $L(N)$ is consistent with a linear function of $N$.
On the other hand, for the rural street network in the UK, we find a different behaviour statistically significant for the same quantity (\textit{p-value=0.007}), which scales in a sub-linear way, i.e. $L(N)\propto N^{0.93}$.
The linear relation for $L$ in urban networks is due to the independence of $\langle k\rangle$ and $\langle l\rangle$ by $N$, while the sub-linear relation for $L$ in the rural network is due to the sub-linearity of $\langle l(N)\rangle$ for those networks (see SI, Sec. III).
 
In the case of cities in California (Fig. \ref{fig4}, panel \textit{b}), we find that the behaviour of $L(N)$ is consistent with that of the UK in a slightly super-linear regime, i.e., $L(N)\propto N^{1.04}$.
On the other hand, for the rural street network in California, we find that $L(N)$  is sub-linear, i.e. $L(N)\propto N^{0.84}$, and it is not  consistent within the error range neither with that of the California urban  street network (\textit{p-value=0.0003}) nor with that of the UK rural street network. 

In panels \textit{c} and \textit{d} of Fig. \ref{fig4}, we see that the exponent for urban network areas $A(N)$ in the UK and in California are quite similar, following a very mild super-linear relation, i.e. $A(N)\propto N^{1.03}$. 
On the other hand, super-linearity can be statistically discarded for both exponents for the rural case in the UK (\textit{p-value=0.000004}) and in California (\textit{p-value=0.0004}). 
In addition, it is important to note that for the rural networks, the exponents for the UK and California are notably different. 
Linearity can discarded for California, while this is not the case for the UK.
 
These differences reflect the contrast in the spatial patterns of the street networks covering these two countries. 
In particular, the nearly linear relations found for the urban areas reflect the fact that street intersections are generally homogeneously distributed within the urban fringes.
Such homogeneity can be seen from the street intersection distributions $P(n)$ shown in panels \textit{e} and \textit{f} of the same figure.
In this case again we find very similar patterns between the UK and California, where $P(n)$ is well fitted by a logistic distribution in the case of urban street networks. This is a bell shaped distribution with a well defined average and variance, while it is ill defined for rural street networks. 

The analysis above highlights the fact that urban street networks are characterized by an overall homogeneous texture, which is consistent between the two different countries considered in this work.
In the same way, we can observe how rural street networks differ consistently from urban street networks and  between different countries, displaying an overall inhomogeneous structure. 
Hence, we find that for urban conglomerations a general behaviour emerge in the study of the scaling laws which characterize the global street network structure.

These hints of universal behaviours do not imply that different cities look the same. 
In fact, different urbanization processes shape cities in very different ways, in terms of morphology and size.
Nevertheless, the compatibility between the exponents for the analysed quantities suggests that there might be common principles for the growth of cities.
If this is the case, then cities at a specific point in time represent different states of the evolutionary process. We will then expect to find a similar behaviour if we looked at the evolution in time of a specific city.
%Our next goal is to understand if these quantities also have the same behaviours during the growth of cities, which is to say if the urban street networks display ergodic behaviours. 
In order to test this hypothesis, we consider a unique dataset recording the evolution of street networks of Greater London  between 1786 and 2010, through nine well spaced temporal intervals defined by the maps shown in Fig. \ref{fig5} (see SI Sec. I C for more info). 

In  Fig. \ref{fig6}, we perform a simple test, by measuring the  aforementioned quantities in the contemporary UK urban street networks and in the  historical London dataset. 
%If the scaling exponents were universal, we would find the same scaling behaviour in the time average as in the space average. 
Interestingly enough, for  $L(N)$ in the UK,  the historical dataset overlaps with the spatial dataset and both allometric fittings are consistent over a linear regime. 
As we stated above, this means an overall homogeneity in terms of the average connectivity $\langle k\rangle$ and the average street segment length $\langle l\rangle$, that is preserved over time.
For $A(N)$, even if the points do not really overlap,  the allometric behaviour is consistent between the time and space averages in the slightly super-linear regime.

\section{Discussion}
Two important results can be derived from our study. On the one hand, we provided a methodology to define city boundaries through spatial urban networks in a universal way. 
On the other, we explored the generality of some scaling laws related to urban street networks. 
Both of these aspects relate to the quest for methodological advancements in the analysis of  spatial urban networks, and they relate to the discussion of important statistical phenomena, such as those described by  Zipf's and Gibrat's law.

Regarding the concept of city boundaries, we discovered universal properties of street networks related to clustering properties in the street intersection space.
These properties allow us to distinguish the urban agglomerate with a methodology that is parameter free and that reduces the problem to extract city boundaries to a simple clustering process on a spatial point pattern.

The concept of city boundaries is very important to distinguish between urban and rural networks. 
We show that allometries found in urban street networks consistently differ from the ones found in rural street networks. 
This means that an ill posed definition of boundaries, such as arbitrary administrative boundaries, would mix the properties of street networks that are in two distinct phases of their evolution, producing spurious results (see SI Sec. III for a direct example). 

Regarding our analysis about the generality in space and time of relevant allometries found in urban street networks,  we chose two very distinct datasets, that present different urbanisation paths.
While  cities in the UK are mostly of Roman or Medieval origin and reflect a long line of urban evolution spanning two millennia,  cities in California are mostly the result of an urban explosion during the latter half of the nineteenth and the twentieth centuries.  
In this context, we find that urban street networks display compatible properties, even though the datasets are very different.
This highlights how the city is an overall homogeneous structure in terms of its street network quantities (average degree, average street length, etc.). 
These findings are confirmed by our  analysis, which compares the structure of the urban street networks in the UK with the street networks of the historical evolution of London during more than two centuries. Even if these results are not definitive, a general behaviour for the found exponents cannot be excluded at this point and new perspectives of research in this direction are thus opened.

Spatial networks are widespread in nature and it is possible to see how the organization of spatially embedded structures are often similar for a variety of different phenomena. 
Leaf venation, crack pattern formation, river networks, ant galleries, circulatory systems, soap froths,  pipe networks and so on, have been studied in a wide range of  disciplines which are often strongly related \cite{westnat,banavar,ants,bohn2007,klein}. 
In particular, brain networks seem to  share a number of similarities with the organization of spatial street networks, due to their high  modularity and fractal structure \cite{gallos2012small}.

Even though cities present a diverse range of morphological features, we have shown that the boundaries of cities can be identified through universal properties of street networks. 
This opens up new research perspectives in terms of the analysis of the  logistic parameters for each city. 
As cities undergo  different stages of evolution, related either to expansion and to condensation phases, those different evolution phases could be easily recognised and classified from the deviations in the logistic curve related to the clustering process (see SI, Sec. II-A).

Moreover, from our analysis we can derive a broad picture of the way a city evolves.
What we observe is that the street network can be found in two very distinct phases, the rural one, which is not characterized by any distinctive properties, and the urban one which is characterized by high density of intersections which are distributed in patterns that are mostly homogeneous and which carry very similar exponents. 
In such a picture a city street network develops as an articulated organism territorializing the sparse rural street network, filling the space with denser residential patterns and then radically changing its morphology.

A key advantage of our method of analysis, compared to other existing approaches, such as those based on data extracted from satellite imagery, is the ease of use. Recent advances in GIS technologies have led to the proliferation of street network data generated by public and private entities. Our study demonstrates that these datasets can be deployed in new ways to analyse key properties of cities, enhancing our ability to manage the built environment.
A disadvantage of our methodology, as it is presented in this form, derives from its bottom-up approach.
As a matter of fact, it is especially indicated to extract a limited number of cities, as the extraction procedure could not be completely automated and needs eye inspection (see SI Sec. II A). 
In order to extract a large number of cities top-down techniques, such as the one presented in \cite{elsa} are definitively more efficient, even if less precise. 

\section*{Acknowledgements}
APM was partially funded by the Engineering and Physical Sciences Research Council (EPSRC) SCALE project (EP/G057737/1), EA, and MB by the European Research Council (ERC) MECHANICITY Project (249393 ERC-2009-AdG), and KS by the ESRC TALISMAN Project (ES/I025634/1). We acknowledge helpful discussions with Dr. Roberto Murcio.

\section*{Author contributions statement}
 A.P.M conceived the study,   A.P.M, E.A. and M.B. conducted the study, A.P.M, E.A. and M.B. analysed the results, A.P.M,  E.H wrote the algorithms, K.S. extracted the historical dataset.  All authors reviewed the manuscript.

\end{document}